\documentclass[journal=jacsat,manuscript=article]{achemso}

\usepackage[version=3]{mhchem} % Formula subscripts using \ce{}
\usepackage{amssymb}
\usepackage{svg}
\author{Tzu-chen Liu}
\affiliation{Department of Materials Science and Engineering, Northwestern University}

\author{Chris Wolverton}
 \email{c-wolverton@northwestern.edu}
\affiliation{Department of Materials Science and Engineering, Northwestern University}

\title{Long- and Short-Range Anion Order in SrTiO$_{3-x}$H$_x$ Perovskite Oxyhydrides: \\
DFT+$U$ Sensitivity and HSE06 Cluster Expansion}

\abbreviations{IR,NMR,UV}
\keywords{American Chemical Society, \LaTeX}

\begin{document}

%\begin{tocentry}

%\end{tocentry}

\begin{center}
(\date{\today})    
\end{center}

\begin{abstract}
Anion ordering in perovskite oxyhydrides can remain significant even in disordered states, particularly at non-dilute hydrogen concentrations.
Nevertheless, hydride substitution poses challenges for accurate simulations based on density functional theory due to configurational complexity and Ti 3$d$ reduction.
Here, we develop a cluster expansion (CE) framework for SrTiO$_{3-x}$H$_x$ incorporating HSE06 hybrid-DFT energetics.
We first demonstrate that calculated mixing energies and ordering stability are highly sensitive to the choice of DFT+$U$, with maximum variations on the order of 100 meV/anion.
We benchmark ordering energetics against HSE06 calculations and identify $U$ = 3.3 eV as an HSE06 proxy, which enables extensive configurational exploration while limiting costly HSE06 calculations to key configurations for efficient learning of ordering energetics.
Together, ground-state orderings, correlations between octahedral configurations and structural stability, and MC sampling of CE models all support a strong preference for the O$_4$H$_2$ cis configuration in SrTiO$_{3-x}$H$_x$, in which two hydride ions occupy first-nearest-neighbor anion sites.
This cis-type preference was overlooked in previous ATiO$_{3-x}$H$_x$ studies, despite its sizable stabilization of ~200 meV per hydride comparable to reported anion-migration and polaron-formation energies.
This study addresses both the previously underexplored sensitivity of CE-based ordering analyses to DFT+$U$ and anion-ordering in perovskite oxyhydrides.
\end{abstract}

\clearpage

\section{Introduction}
Incorporating hydride ions into the oxide lattice of perovskite ATiO$_{3-x}$H$_x$ (A= Ba, Sr, and Ca)\cite{kobayashi2012oxyhydride,sakaguchi2012oxyhydrides} yields a rare class of oxyhydrides that exhibit functionalities for catalysis, electrochemistry, and hydrogen science applications.\cite{kobayashi2017titanium, kageyama2018expanding, tang2018metal, liu2019highly,granhed2019band,fine2025unraveling}
In ATiO$_{3-x}$H$_x$, the hydride content has been reported to reach non-dilute levels of $x=0.6$, depending on the A-site species.
More generally, anion ordering has been observed in perovskite structures\cite{kageyama2018expanding, yang2011anion}, with its strength and impact expected to become more pronounced at anion substitution levels beyond the dilute regime.
Two different anion species inevitably exhibit some ordering preference unless the system approaches the ideal-solution limit, which is unlikely for the aliovalent substitution.
Even for those oxyhydrides reported as disordered at finite temperature, materials are known to exhibit short-range order (SRO) that deviates from randomness.\cite{cowley1950approximate,cowley1960short}
However, prior computational modeling of ATiO$_{3-x}$H$_x$ commonly consider more dilute and potentially oversimplified pictures assuming hydride ion (H$^-$) isolated; for example, a single H$^-$ in a $2\times2\times2$ or $3\times3\times3$ supercell.\cite{liu2019highly,granhed2019band,fine2025unraveling}
Studies of ATiO$_{3-x}$H$_x$ still lack a comprehensive understanding of anion ordering behavior at non-dilute compositions, a key factor that influences material properties in both simulations and experimental measurements.\cite{pilania2020anion,kageyama2018expanding}

\begin{figure}
\centering
\includegraphics[width=\columnwidth]{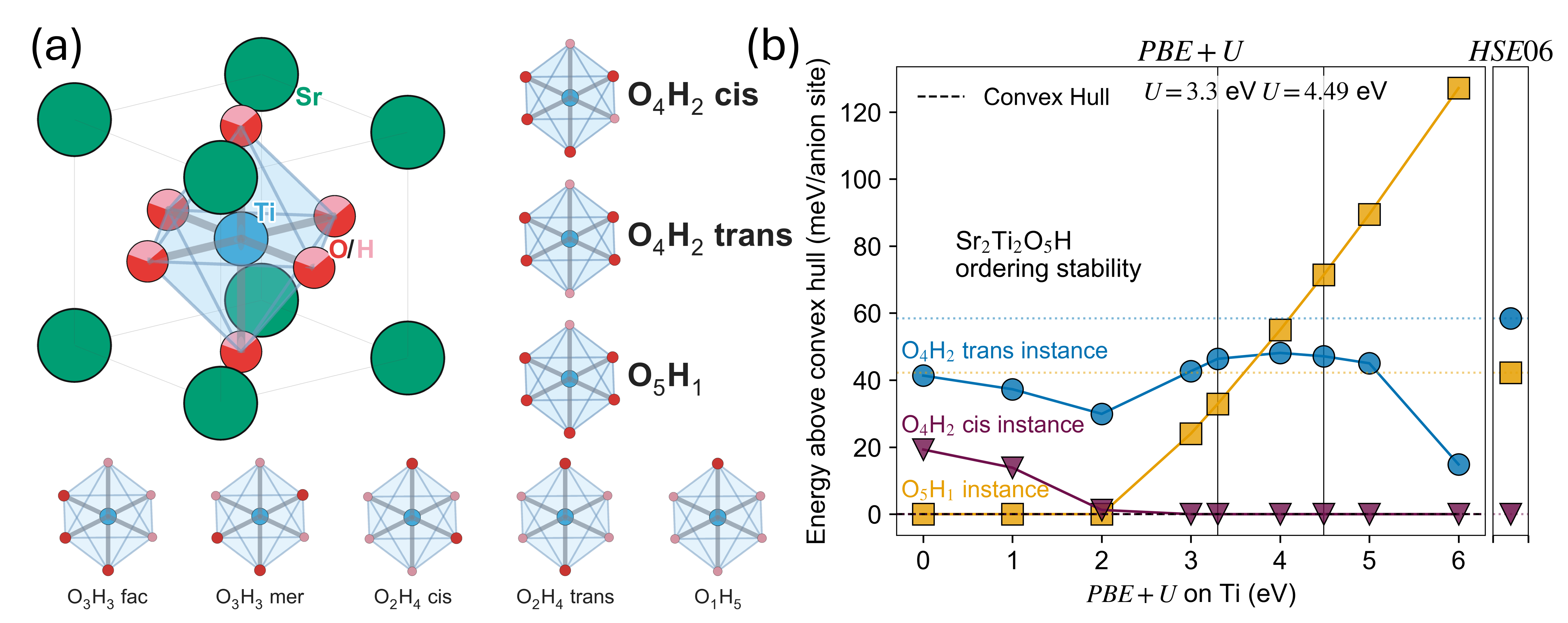}
\caption{\label{fig:intro} Anion configurational complexity and the associated DFT+$U$ sensitivity.
(a) Crystal structure of SrTiO$_{3-x}$H$_x$ and possible anion mixing configurations of O$_{6-x}$H$_x$ octahedra surrounding Ti centers. 
In the oxyhydride region $x \leq 1$ considered in this work, the primary competing octahedral configurations are O$_4$H$_2$ cis, O$_4$H$_2$ trans, and O$_5$H$_1$.
(b) Sensitivity of DFT+$U$ ordering stability, demonstrated using three SrTiO$_5$H structures in which hydrogen substitutions lead to O$_4$H$_2$ cis, O$_4$H$_2$ trans, and O$_5$H$_1$ configurations. 
Note that these three structures are neither unique nor representative of each configuration class, and are shown only for conceptual illustration. 
The formal ordering energetics analyses are presented in the Results and Discussion section.}
\end{figure}

The accurate simulation of anion order in ATiO$_{3-x}$H$_x$ (in this work, SrTiO$_{3-x}$H$_x$) using density functional theory (DFT)\cite{hohenberg1964inhomogeneous, kohn1965self} is challenging because H$^-$ substitution simultaneously introduces anion configurational complexity (Figure \ref{fig:intro}a) and Ti-3$d$ reduction.
The cluster expansion (CE) method is well established for treating configurational complexity.\cite{sanchez1984generalized, lu1991first, de1994cluster, van2002automating}
When constructing a CE model, accurate DFT simulations to obtain small energetic differences (on the meV/atom scale) across configurations are necessary.
Nevertheless, DFT at the local density approximation (LDA) and generalized gradient approximation (GGA) semilocal levels is known to over-delocalize electrons, commonly associated with self-interaction error.\cite{perdew1981self, anisimov1997first, Martin_2004, zhou2004first, perdew2009some, himmetoglu2014hubbard, kulik2015perspective}
This over-delocalization tendency can be especially problematic for localized transition metal $d$ states on reduced Ti cations in systems such as ATiO$_{3-x}$H$_x$, perovskites containing oxygen vacancies, and other 3$d^1$ Ti-containing compounds.\cite{pavarini2004mott, cuong2007oxygen, nathan2024peierls}
To remediate over-delocalization, DFT+$U$ has been a popular scheme in solid state materials community that provides improved electron descriptions with modest additional cost.\cite{anisimov1991band, liechtenstein1995density, dudarev1998electron} 
This approach is widely used in resource-intensive tasks such as high-throughput DFT databases and training CE models or modern machine-learned interatomic potentials (MLIPs).\cite{ saal2013materials, kirklin2015open, jain2013commentary, curtarolo2012aflow}
However, we show in this work that the choice of U value can strongly affect assessments of ordering energetics and ground states (Figure \ref{fig:intro}b), and therefore necessitates careful considerations.

DFT+$U$ can suffer from limited transferability of $U$ values, whereas CE construction requires consistent energies and therefore limits the use of varying $U$ values across training structures of the same model.\cite{aykol2014local, capdevila2016performance, jain2011formation, liu2025anomalous, warford2026better}
Prior CE studies for systems with DFT over-delocalization commonly either use no $U$ or apply a single $U$ value.\cite{wolverton1998cation, urban2016computational, ji2019hidden, ouyang2019cluster, zhang2024screening}
Such an approach may be insufficiently physically accurate for anion ordering in ATiO$_{3-x}$H$_x$, but it is also difficult to find an improvement within the DFT+$U$ scheme, as discussed in the following arguments: 
(1) Even for a single compound, a single $U$ value generally cannot produce accurate predictions of many properties (e.g., oxidation energies, lattice parameters, band gaps, etc.).\cite{capdevila2016performance, liu2025anomalous}
Given this limited transferability, there is no clear way to choose a single $U$ value that is optimal for accurately predicting ordering stability (i.e., the energetic competition among many different ordered structures).
(2) For a single element, different oxidation states and ligands might require different strengths of $U$ corrections.\cite{aykol2014local}
Such variability complicates the construction of a CE for ATiO$_{3-x}$H$_x$, where configurations exhibit different levels of H$^-$ incorporation leading to varying oxidation states and degrees of electron localization. 
This heterogeneity is even present locally in each structure, since each Ti can be coordinated by different numbers of O and H within the coordinating O$_x$H$_{6-x}$ octahedra.
(3) Finally, constructing a CE model requires energetic comparisons among configurations. 
Even if one could determine distinct $U$ values for each calculation, a correction scheme to evaluate energies consistently would be prohibitive in CE construction.\cite{jain2011formation, aykol2014local}
Compared with conventional energy convex-hull constructions, CE focuses on dense variations in compositions and configurations associated with meV/atom-level energetic differences, a scale at which experimental values are generally not available to set the corrections.
These concerns and limitations surrounding DFT+$U$ motivate the development of a new scheme based on a different physical treatment of electron over-delocalization in DFT.

An alternative way to overcome over-delocalization in semilocal DFT is to climb ``Jacob’s ladder'' of exchange-correlation functionals to achieve higher chemical accuracy.\cite{perdew2009some, perdew2001jacob}
We first test the meta-GGA level (in this work, r$^2$SCAN)\cite{furness2020accurate} and find that its predictions of ordering stability, including the trend with increasing $U$ and corresponding transitions, are similar to those at the GGA level. %as discussed in more detail in the Results section.
We then turn to the next rung of Jacob’s ladder, hybrid functionals (in this work, Heyd-Scuseria-Ernzerhof (HSE06)\cite{heyd2003hybrid, krukau2006influence} range-separated hybrid functional calculations).
Hybrid functionals are commonly considered to exhibit higher chemical accuracy by including a portion of Hartree-Fock (HF) exact exchange, which indeed cancels out the self-interaction in the Hartree term as seen in the original HF formulation.\cite{perdew2001jacob,Martin_2004,kulik2015perspective,chevrier2010hybrid}
Moreover, hybrid functionals have shown success in predicting the structural, electronic, and vibrational properties of Ti-containing perovskites.\cite{wahl2008srtio, janotti2011strain, el2011modeling, el2013neutral, el2013structural, granhed2019band}
Nevertheless, we note that there is currently no direct evidence that hybrid functionals necessarily yield the most accurate ordering stability in ATiO$_{3-x}$H$_x$.
Hybrid functionals may also exhibit a risk of over-localization error from their HF component.\cite{mori2008localization, cohen2008insights, granhed2019band}
Despite these caveats, we consider HSE06, given its reputation for high chemical accuracy, success in Ti-containing perovskites, and the consistent Hamiltonian for mitigating over-delocalization,\cite{chevrier2010hybrid} to be a reasonable choice for studying the anion ordering in perovskite oxyhydrides.
However, a final practical consideration is the high computational cost of hybrid functionals, especially in the plane-wave basis used in most periodic DFT,\cite{chevrier2010hybrid, kulik2015perspective, carnimeo2019fast} which commonly exceeds the cost of semilocal DFT by one or even two orders of magnitude.
This cost is further exacerbated in this study since CE constructions for complex systems might require hundreds of training structures to meet the standard CE quality criteria.\cite{van2002alloy}
Simultaneously accounting for anion configurational complexity and over-delocalization in DFT energetics presents a substantial technical challenge, which we aim to address as part of this work.

\begin{figure}
\centering
\includegraphics[width=\columnwidth]{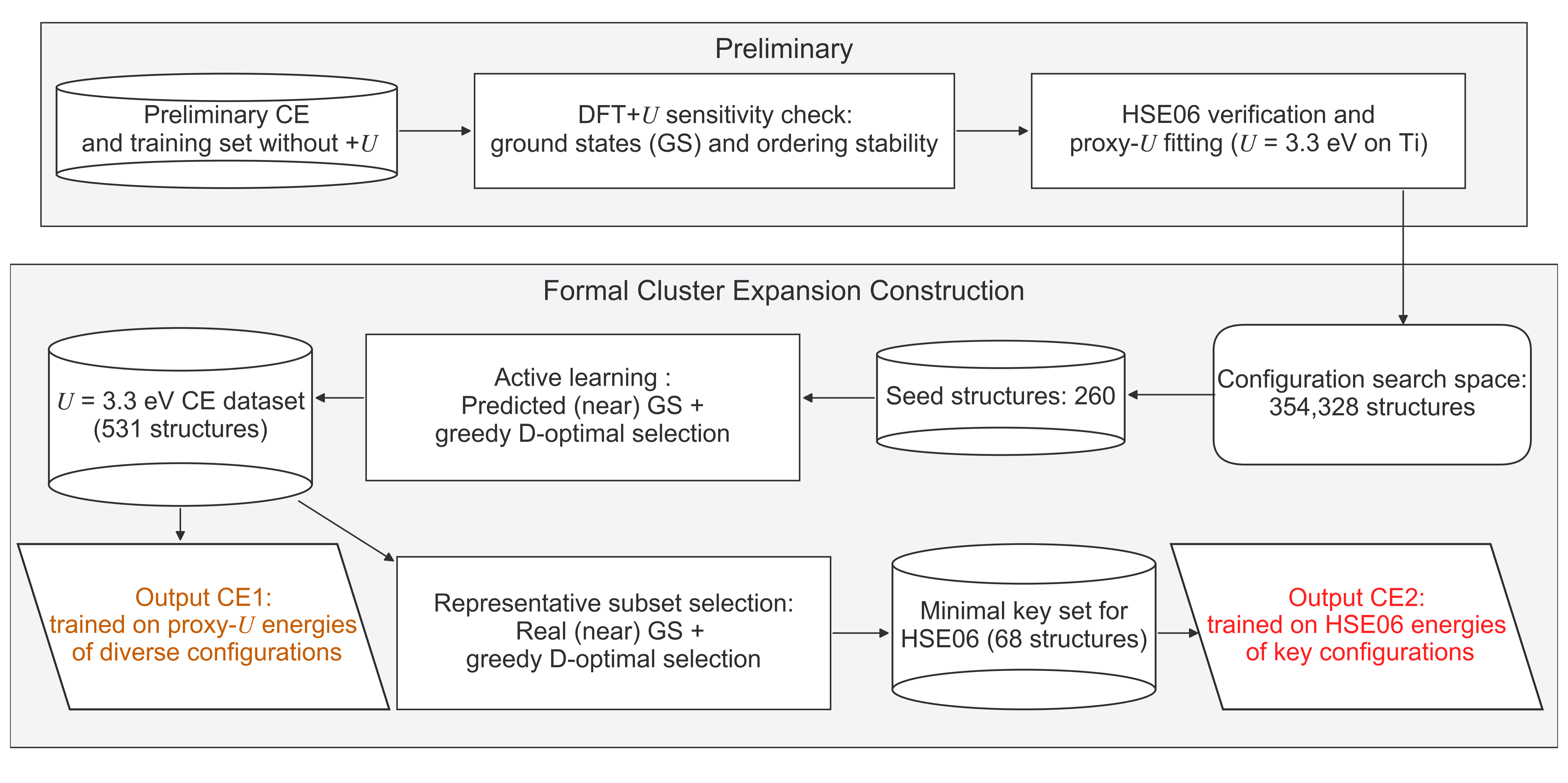}
\caption{\label{fig:framework} Workflow for resolving DFT+$U$ sensitivity and constructing converged CE with HSE06 energetics. 
This framework first yields an intermediate output CE1 trained on the ordering energetics of 531 structures with diverse configurations, selected by active learning from a configuration space of 354,326 structures, with DFT+$U$ ($U$ = 3.3 eV on Ti, also denoted as proxy-$U$ where convenient) used as an HSE06 proxy determined in the preliminary stage. Representative subset selection is then performed to identify key configurations for the HSE06 CE fitting, yielding the final output CE2 trained with HSE06 energetics.     
}
\end{figure}

In this work, we construct a CE model trained on energies from HSE06 to investigate anion order in SrTiO$_{3-x}$H$_x$.
Since HSE06 calculations are computationally demanding, we propose a new HSE06 CE training framework (Figure \ref{fig:framework}) that generates representative structures to minimize the number of HSE06 calculations required for the converged CE construction. 
Still, the framework sufficiently sampling the wide configurational search space and capturing ground-state orderings across 354,328 possible configurations in supercells of up to 40 atoms.
The framework combines a DFT+$U$-to-HSE06 fitting of ordering stability, an active-learning scheme, and greedy D-optimal selection for the HSE06 CE training set. 
DFT+$U$ ordering stability, a sensitive function of $U$ values, is compared with HSE06 ordering stability to determine $U$ = 3.3 eV (hereafter denoted as proxy-$U$) serving as a low-cost proxy of HSE06 for structure-selection steps of CE construction. 
CE results indicate a prevalent cis-type ordering preference in lower-energy configurations across the composition range of $x$ = 0 to 1 in SrTiO$_{3-x}$H$_x$. 
To demonstrate the relevance of anion ordering even in the dilute limit, we compare an isolated H with two H placed in a single cis- and trans-type configuration to calculate their binding energies. 
We found that cis-type binding is about 200 meV/H more stable than isolated H, an energy scale comparable to, or even larger than, other energetic quantities such as migration barriers and polaron formation energies reported in prior literature.\cite{granhed2019band, liu2019highly}
Our work advances previously underexplored aspects of anion ordering tendencies in perovskite oxyhydrides and DFT+$U$ sensitivity in CE for transition metal oxides.
The framework for constructing HSE06 CE models could potentially be extended to similar research problems for consistent modeling of systems with varying degrees of electron (de)localization.

\section{Methodology}
\subsection{DFT Calculations}
Spin-polarized DFT calculations were computed using the Vienna Ab initio Simulation Package (VASP).\cite{kresse1993ab, kresse1996efficient, kresse1996efficiency} The projector augmented-wave (PAW) \cite{blochl1994projector, kresse1999ultrasoft} pseudopotentials were used, with the following states treated as valence: $4s^{2}4p^{6}4d^{0.001}5s^{1.999}$ for Sr, $3s^{2}3p^{6}3d^{3}4s^{1}$ for Ti, $2s^{2}2p^{4}$ for O, and $1s^{1}$ for H (namely, the VASP POTCAR Sr\_sv, Ti\_sv, O, and H).~\cite{VASP_psuedo}
The exchange-correlation functionals PBE,\cite{perdew1996generalized} r$^2$SCAN, and HSE06 were used for the GGA, meta-GGA, and hybrid (hyper-GGA) calculations, respectively.\cite{perdew2009some, perdew2001jacob}
HSE06 calculations were performed with the default values of the mixing (0.25) and range-separation parameter (0.2 $\text{\AA}$$^{-1}$).
DFT+$U$ for both PBE and r$^2$SCAN were performed using the simplified rotationally invariant scheme from Dudarev et al.,\cite{dudarev1998electron} with effective $U$ values on Ti varied from 0 to 6 eV to examine how Hubbard corrections impact ordering stability and ground states. 
For all calculations, the plane-wave basis set with 550 eV cutoff energy was used. 
The k-point meshes for the Brillouin zone sampling were generated with k-point spacing along each reciprocal lattice vector direction (KSPACING) of 0.3 $\text{\AA}$$^{-1}$ for coarse structural relaxations and 0.2 $\text{\AA}$$^{-1}$ for both fine structural relaxations and final static runs.
Both the energy cutoff and k-point meshes were tested to ensure convergence of ordering stability within 1 meV/atom relative to stricter settings of 680 eV cutoff energy and KSPACING of 0.3/0.2/0.15 $\text{\AA}$$^{-1}$ for coarse relaxations/fine relaxations/static runs, respectively, on a selected subset of structures.
Initial magnetic moments on Ti were assigned in ferromagnetic configurations, while nonmagnetic and antiferromagnetic configurations were tested and found not to lower the energy.

Due to the high accuracy required for ordering stability investigations and CE construction, working at the meV/anion (mixing site) scale, we perform structural relaxations at the HSE06 level to obtain HSE06 equilibrium geometries and energies.
We found that performing HSE06 only as a static calculation on semilocal DFT (with or without +$U$) relaxed geometries yields energies higher by about 3 to 15 meV/atom with overestimated volumes by roughly 2 to 5\% relative to HSE06 relaxed results, and these deviations can vary non-uniformly across configurations and thus affect the predicted ordering stability.
We note that PBE or r$^2$SCAN relaxations without +$U$ give volumes closer to the HSE06 equilibrium volume, while, as shown in the result section, $U$ = 3.3 eV on Ti yields ordering stability quantities closer to HSE06 results. 
To avoid geometry-induced concerns and bias toward a particular $U$ when selecting a proxy for HSE06, we consider HSE06 calculations requiring their own structural relaxations in this study.

For DFT+$U$-to-HSE06 fitting of ordering stability and ground-state verifications, where we focused on a selected set of carefully validated calculations, structural relaxations in the HSE06, r$^2$SCAN, and PBE calculations were converged such that all force norms were below 0.01 eV/$\text{\AA}$, with 10$^{-6}$ eV for electronic-loop energy convergence, using the fine relaxation settings mentioned above.
Proxy-$U$ calculations in CE training structure selection stages were performed with the sequence and settings of coarse relaxations/fine relaxations/static runs mentioned above. 
HSE06 CE training structures (not already included in the previous DFT+$U$-to-HSE06 mapping set, mostly larger supercells, for which their HSE06 relaxations are particularly computationally demanding), were first relaxed with a coarse criteria 0.05 eV/$\text{\AA}$ for forces and 10$^{-4}$ eV for electronic loop. 
These relaxed structures were then followed by the final HSE06 static calculations with the 10$^{-6}$ eV energy convergence. 
This practice is found to produce final energies that only differ by around 1 meV/atom compared with calculations converged to 0.01 eV/$\text{\AA}$ in forces in verified cases.

\subsection{Cluster Expansion and Monte Carlo Sampling}

Cluster Expansion is a standard formalism to represent the configurational degrees of freedom using a typically orthonormal cluster basis:

\begin{equation}
    Q(\sigma)=\sum_{\alpha} m_{\alpha} J_{\alpha} \langle \Gamma_{\alpha}(\sigma)\rangle
   \label{eq:Method_CE} 
\end{equation}
where $Q(\sigma)$ is the quantity of interest as a function of configuration $\sigma$; $\alpha$ denotes each symmetrically inequivalent cluster; $m_\alpha$ is the symmetry multiplicity of $\alpha$; $J_\alpha$ is the expansion coefficient for $Q$, commonly called the effective cluster interaction (ECI) when used in a mixing-energy expansion; and $\langle \Gamma_{\alpha}(\sigma)\rangle$ is the correlation function of each cluster, calculated as the average (over clusters symmetrically equivalent to $\alpha$) of the product of the occupation variables ($\pm 1$ for the binary case of O/H anion sublattice mixing) for sites $i$ in $\alpha$.

The CE framework in this work is utilized both to efficiently explore the (O/H anion sublattice) configurational space of SrTiO$_{3-x}$H$_x$ and to generate a Hamiltonian for MC sampling of finite-temperature properties. 
The Alloy Theoretic Automated Toolkit (ATAT) \cite{van2002alloy,van2002automating} was used in the initialization stage for configuration generation and ground-state searches within its standard automatic CE construction workflow using PBE without +$U$. 
The generated anion ordering configurations and no-$U$ ground state structures were then used in the preliminary stage in Figure \ref{fig:framework} for DFT+$U$-to-HSE06 fitting.
The HSE06 CE framework incorporates CE functions from the integrated cluster expansion toolkit (ICET) \cite{aangqvist2019icet} Python package.
The set of 354,328 configurations with supercells up to 40 atoms/supercell for $x < 1$ compositions and up to 30 atoms/supercell for $x=1$ (O$_2$H) (to prevent the training set from being overwhelmingly dominated by the large enumerations at this composition) was designated as the search space.
The training structures were then selected from this space through D-optimal design and (near) ground-state search in the formal CE construction stage shown in Figure \ref{fig:framework}. 
The 260 seed structures were similarly selected from supercells containing up to 20 atoms/supercell for $x \le 1$ compositions and up to 15 atoms/supercell for $x = 1$.
We do not consider compositions with $x>1$ since perovskite oxyhydrides are rarely reported to exceed an O$_2$H ratio, and thus we exclude this regime in this study and focus on $x \le 1$ for SrTiO$_{3-x}$H$_x$.

The active-learning scheme proceeds as follows: an initial CE model is trained on the seed structures, and the resulting Hamiltonian is then used to predict the energies of 354,328 configurations and identify new ground-state candidates, which are included in the next round of DFT calculations.
As the ground-state predictions began to approach the following criteria: (1) the true and CE-predicted ground states agree within the DFT-calculated set and (2) no configuration within the set of 354,328 structures is predicted to be more stable than the current convex hull, additional configurations selected by D-optimal design were added to improve the spanning of correlation space.
Together with above ground-state criteria, the training process was considered complete once (3) the CE model reached a satisfactory CV score and (4) showed converged ECIs, with interaction magnitudes decaying to near zero before the cutoff so that longer-range clusters could be reasonably neglected.\cite{van2002alloy}

The CE model fitting steps in Figure \ref{fig:framework} were all performed using adaptive LASSO with hyperparameter search for cluster cutoffs, the regularization parameter of the adaptive LASSO, and importance weights on structures.
During CE fitting, assigning weights on ground-state and near-ground-state configurations sacrifices cross-validation (CV) performance but improves ground-state reproduction and helps maintain physical correctness of the model. %in subsequent MC sampling for anion order predictions at finite temperature.
In the active learning stage using PBE+$U$, where training structures are abundant and CE fitting is straightforward, only the regularization parameter is scanned, while the cluster cutoffs and weights are assigned manually.
In the final HSE06 CE fitting with a minimal training set, hyperparameter optimization using Optuna was performed to tune the regularization parameter, cluster cutoffs, and weights simultaneously.\cite{optuna_2019} 
The optimization was divided into two stages: global exploration with a Tree-structured Parzen Estimator (TPE)\cite{bergstra2011algorithms}, followed by local refinement with Covariance Matrix Adaptation Evolution Strategy (CMA-ES)\cite{hansen2016cma}.
Weights can be assigned to all training structures, but were specifically increased for configurations on or near the convex hull.
This practice balances physical correctness and CV performance under the challenging fitting conditions imposed by the small HSE06 training set.

\begin{figure}
\centering
\includegraphics[width=0.8\columnwidth]{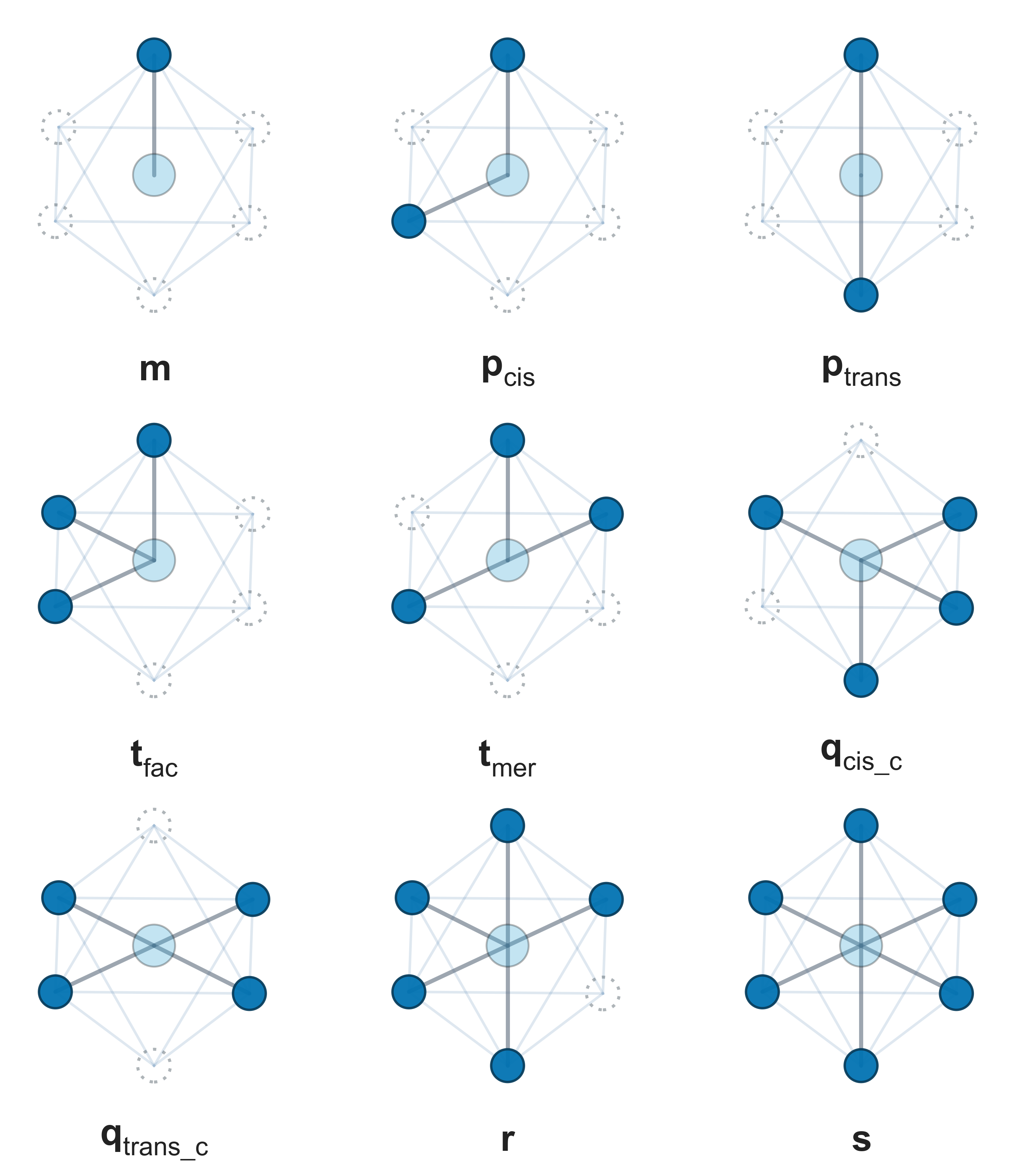}
\caption{\label{fig:cluster} 
Illustration of the clusters included in Equation \ref{C_matrix}. 
The sites included in each cluster are selected from the corners of the Ti-containing octahedra and are shown as dark-blue spheres, with Ti represented by the light sky-blue sphere at the center. 
The remaining corners, which are not part of the cluster, are indicated by dotted circles. 
Among the clusters shown here, those with two symmetry-inequivalent geometries within the same order ($p$, $t$, $q$, or $s$) are further distinguished.
Since geometries of cluster sites share features similar to those defined for the O/H configurations in Figure \ref{fig:intro}a, the corresponding subscript labels are adopted from that notation.
The ``complement (\_c)'' naming for the quadruplets is given because the remaining unselected sites in octahedra form a trans or cis geometry.
}
\end{figure}

The MC simulation was conducted using the emc2 function in ATAT to sample anion ordering quantities in disordered states.
Reported ensemble averaged quantities were obtained using a simulation cell of 26$^3$ sites, with 20,000 MC flips per site for equilibration and 40,000 MC flips per site for averaging.
The octahedral probabilities and cluster correlations obtained from MC sampling can be connected through the configuration matrix (C-matrix) formulation \cite{ceder1991alloy,de1992cluster,liu2026short}:
\begin{equation}
\label{C_matrix}
\begin{pmatrix}
P(\mathrm{O}_{6})\\
P(\mathrm{O}_{5}\mathrm{H}_{1})\\
P(\mathrm{O}_{4}\mathrm{H}_{2}\,\mathrm{cis})\\
P(\mathrm{O}_{4}\mathrm{H}_{2}\,\mathrm{trans})\\
P(\mathrm{O}_{3}\mathrm{H}_{3}\,\mathrm{fac})\\
P(\mathrm{O}_{3}\mathrm{H}_{3}\,\mathrm{mer})\\
P(\mathrm{O}_{2}\mathrm{H}_{4}\,\mathrm{cis})\\
P(\mathrm{O}_{2}\mathrm{H}_{4}\,\mathrm{trans})\\
P(\mathrm{O}_{1}\mathrm{H}_{5})\\
P(\mathrm{H}_{6})
\end{pmatrix}
=
\frac{1}{64}
\begin{pmatrix}
1 & 6 & 12 & 3 & 8 & 12 & 12 & 3 & 6 & 1\\
6 & 24 & 24 & 6 & 0 & 0 & -24 & -6 & -24 & -6\\
12 & 24 & 0 & -12 & 0 & -48 & 0 & -12 & 24 & 12\\
3 & 6 & -12 & 9 & -24 & 12 & -12 & 9 & 6 & 3\\
8 & 0 & 0 & -24 & 0 & 0 & 0 & 24 & 0 & -8\\
12 & 0 & -48 & 12 & 0 & 0 & 48 & -12 & 0 & -12\\
12 & -24 & 0 & -12 & 0 & 48 & 0 & -12 & -24 & 12\\
3 & -6 & -12 & 9 & 24 & -12 & -12 & 9 & -6 & 3\\
6 & -24 & 24 & 6 & 0 & 0 & -24 & -6 & 24 & -6\\
1 & -6 & 12 & 3 & -8 & -12 & 12 & 3 & -6 & 1
\end{pmatrix}
\begin{pmatrix}
1\\
m\\
p_{\mathrm{cis}}\\
p_{\mathrm{trans}}\\
t_{\mathrm{fac}}\\
t_{\mathrm{mer}}\\
q_{\mathrm{cis\_c}}\\
q_{\mathrm{trans\_c}}\\
r\\
s
\end{pmatrix}
\end{equation}
where the right-hand side column lists the correlations of clusters shown in Figure \ref{fig:cluster}: $m$ for the point cluster, $p$ for pair clusters on cis and trans sites, $t$ for triplets on fac and mer sites, $q$ for quadruplets on cis-complement (cis\_c) and trans-complement (tans\_c) sites, $r$ for the quintuplet, and $s$ for the sextuplet. 
The C-matrix above is derived using correlations calculated by assigning occupation variables of -1 and 1 to H and O, respectively.
Please note that all clusters in this formulation are subclusters of O$_{6-x}$H$_x$ octahedra surrounding Ti centers; clusters including trans sites but without Ti in between are not included. 
The octahedral probabilities reported in this work are obtained by translating the ATAT output cluster correlations using this equation, and the formula in Equation \ref{C_matrix} was double-checked in several cases by direct counting of octahedra in MC snapshots, yielding consistent results.

%%%%%%%%%%%%%%%%%%%%%%%%%%%%%%%%%%%%%%%%%%%%%%%%%%%%%%%%%%%%%%%%%%%%%%%%%%
\section{Results and discussion}
%%%%%%%%%%%%%%%%%%%%%%%%%%%%%%%%%%%%%%%%%%%%%%%%%%%%%%%%%%%%%%%%%%%%%%%%%%
\subsection{DFT+$U$ Sensitivity and HSE06 Proxy}
%%%%%%%%%%%%%%%%%%%%%%%%%%%%%%%%%%%%%%%%%%%%%%%%%%%%%%%%%%%%%%%%%%%%%%%%%%

\begin{figure}
\centering
\includegraphics[width=0.98\columnwidth]{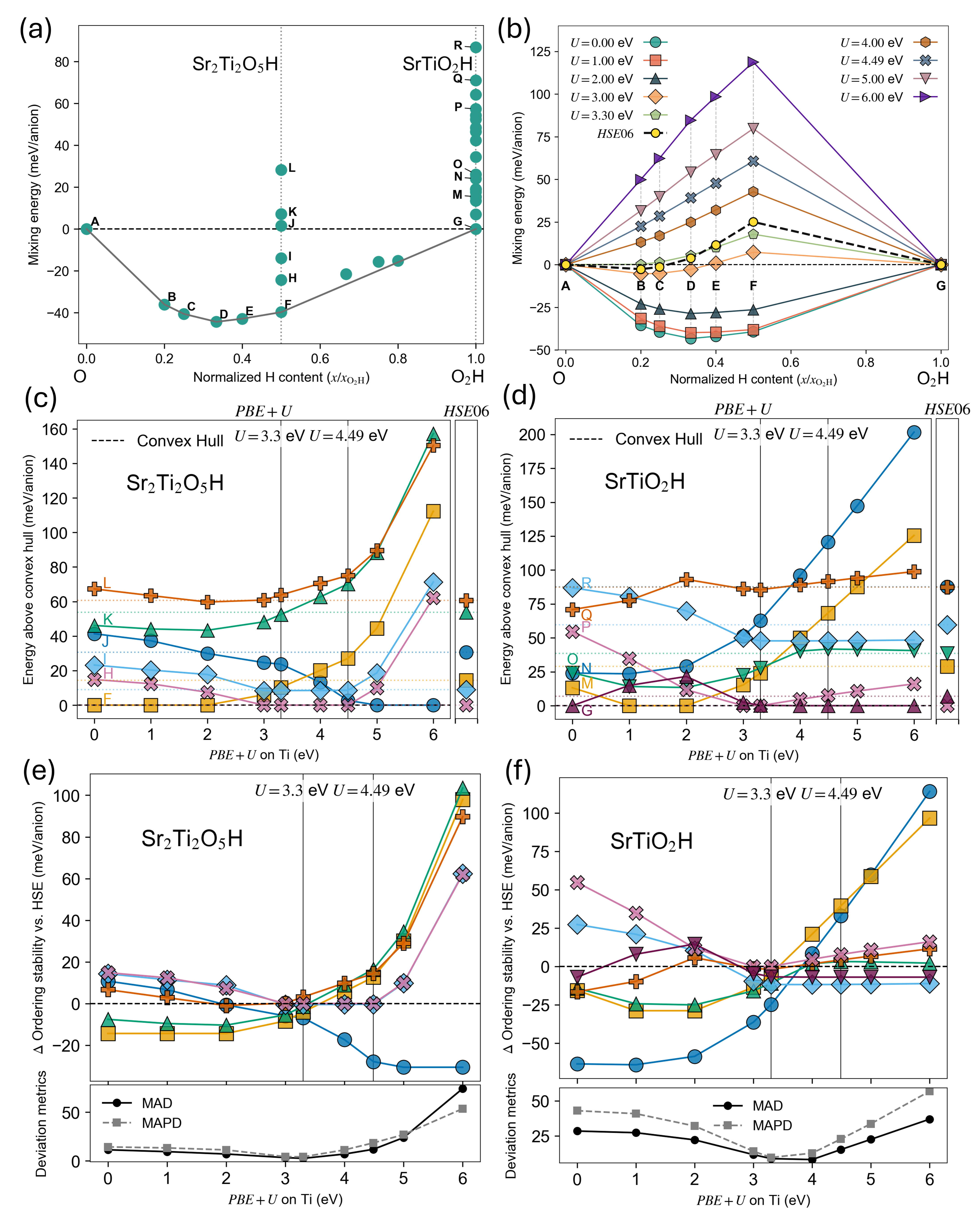}
\caption{\label{fig:sensitivity} PBE+$U$ Sensitivity on ordering stability and HSE06 proxy derivation.
(a) Convex hull construction over the composition range from O to O$_{2}$H (with the lowest-energy O$_{2}$H ordering as the H-rich side reference) constructed by the default CE approach using ATAT and plain PBE without +$U$. (Continued on next page)
}
\end{figure}

\begin{figure}
  \ContinuedFloat
  \caption{(cont.) The alphabet labels denote the orderings that are further examined for their PBE+$U$ sensitivity and verified against HSE06 results. 
  (b) The mixing energies of ground-state orderings (A to G, where A and G are fixed as reference states) on the hull as a function of the $U$ values applied on Ti, compared with HSE06 results. 
  The $U$ = 3.3 eV on Ti yields mixing energies closest to the HSE06 results.  
  The vertical dash lines indicate that these mixing energies can vary on the scale of 100 meV/anion across no $U$ to $U$ = 6 eV.
  Please note that these results do not mean that there will be no stable orderings between O and O$_2$H at high $U$.
  As $U$ increased, the B to F orderings might no longer stay on the hull, but other previously unstable orderings can be stabilized, as demonstrated in the following subfigures. 
  (c) Sr$_2$Ti$_2$O$_5$H and (d) SrTiO$_2$H ordering stability as a function of the $U$ values applied on Ti, from 0 to 6 eV with a step size of 1 eV, and additionally two $U$ values, 3.3 and 4.49 eV from ATiO$_{3-x}$H$_x$ literature.\cite{granhed2019band, liu2019highly}
  The corresponding HSE06 ordering stability is plotted as additional vertical box on the right, with extended dot lines to assist visual comparison. 
  (d) Sr$_2$Ti$_2$O$_5$H and (e) SrTiO$_2$H differences in ordering stability between PBE+$U$ and HSE06 calculations, for the MAD and MAPD evaluations (Equation \ref{eq:mad} and \ref{eq:mapd}) to choose the $U$ value suitable as a low-cost proxy of HSE06, which is then used in the next section of the HSE06 CE construction.
    }
%\label{fig:Figure2-cont}
\end{figure}

We initialize the preliminary stage in Figure \ref{fig:framework} by obtaining the DFT convex hull without +$U$ using the default CE approach in ATAT, as shown in Figure \ref{fig:sensitivity}a. 
The ATAT program includes its own structure selection algorithm to efficiently explore ground states and correlation space to reduce prediction errors.\cite{van2002automating} 
The generated convex hull captures stable orderings (labels A to G) on the hull, as well as selected orderings at Sr$_2$Ti$_2$O$_5$H (label F and H to L) and SrTiO$_2$H (label G and M to R).
Nevertheless, as presented in the subsequent subplots of Figure \ref{fig:sensitivity}, the ordering energetics of all these configurations become sensitive to the $U$ value applied once the Hubbard correction is introduced.
Previously stable orderings can be destabilized by more than 100 meV/anion in mixing energy with high $U$ applied (Figure \ref{fig:sensitivity}b), while originally unstable orderings can exhibit lower energies or even approach to the hull (Figure \ref{fig:sensitivity}c,d).
Since the current dataset was generated through ground-state search without +$U$, the true ground-state orderings at high $U$ are likely not included in the present set of ordered structures and will be further explored in the formal CE construction stage of Figure \ref{fig:framework}.

As discussed in the Introduction section, we consider the HSE06 results to be the reference for examining these variations in ordering energetics.
We can observe that $U$ = 3.3 eV generally reproduces the HSE06 ordering energetics most closely, although not uniformly across the investigated composition range, with larger deviations remain for the H-rich mixing energy (label F in Figure \ref{fig:sensitivity}b) and the ordering stability of SrTiO$_2$H (Figure \ref{fig:sensitivity}d,f). 
We note that, for the mixing energy in Figure \ref{fig:sensitivity}b, G is always used as the zero reference, but it is not always the lowest-energy structure at SrTiO$_2$H for high $U$ or in HSE06 results.
To quantitatively determine the suitable $U$ value as an HSE06 proxy, we define two metrics, the Mean Absolute Deviation (MAD) and Mean Absolute Pairwise Deviation (MAPD), to measure deviations in ordering stability between PBE+$U$ and HSE06, as illustrated in Figure \ref{fig:sensitivity}e,f and their formulations given in following equations:
\begin{equation}\label{eq:mad}
\operatorname{MAD}(U) = \frac{1}{n}\sum_{i=1}^{n} \left| E_i^{\mathrm{PBE+U}}(U) - E_i^{\mathrm{HSE06}} \right|,
\end{equation}

\begin{equation}\label{eq:mapd}
\begin{split}
\operatorname{MAPD}(U) = \frac{2}{n(n-1)} \sum_{1 \le i < j \le n}
\left| \left(E_i^{\mathrm{PBE+U}}(U) - E_j^{\mathrm{PBE+U}}(U)\right) - \left(E_i^{\mathrm{HSE06}} - E_j^{\mathrm{HSE06}}\right) \right| \\
= \frac{2}{n(n-1)} \sum_{1 \le i < j \le n}
\left| \left(E_i^{\mathrm{PBE+U}}(U) - E_i^{\mathrm{HSE06}}\right)- \left(E_j^{\mathrm{PBE+U}}(U)  - E_j^{\mathrm{HSE06}}\right) \right|
\end{split}
\end{equation}
MAD directly represents the absolute difference in ordering stability when compared with the lowest-energy ordering at a given $U$, while MAPD captures the consistency of pairwise ordering-stability rankings.
When two orderings have PBE+$U$ deviations from the HSE06 results in the same direction, the MAPD between that pair of orderings will have a smaller penalty, and vice versa. 
At Sr$_2$Ti$_2$O$_5$H, both MAD and MAPD indicate that  $U$ = 3.3 eV is the optimal setting to match the HSE06 ordering stability.
At SrTiO$_2$H, $U$ = 4.0 eV further lowers MAD from the value at $U$ = 3.3 eV.
However, the $U$ = 4.0 eV results cannot reproduce the HSE06 ordering ranking, particularly between M/O/R and N/Q.
This ordering ranking discrepancy is reflected in the higher MAPD value at $U$ = 4.0 eV.
We consider the ranking consistency captured by MAPD to be more important for studying ordering tendencies.
Based on the overall information shown in Figure \ref{fig:sensitivity}, we determine $U$ = 3.3 eV as the setting of proxy-$U$ used in the remaining investigations of this work.
We note that, as argued in the Introduction section, concerns regarding the physical robustness of DFT+$U$ remain in this challenging case.
This concern is partially reflected in the non-uniform deviations between PBE+$U$ and HSE06 calculations across compositions and configurations even with fixed $U$.
Still, we do not consider this proxy-$U$ usage to be motivated by physical accuracy, nor is this derivation a solid framework for a physically accurate $U$ value for general ordering investigations.
We consider proxy-$U$ to be a convenient, low-cost proxy used in the structure selection stage of CE construction using HSE06 calculations, as shown in the next section. 
At the same time, we will continue to track the differences and agreements between proxy-$U$ and HSE06 in the following analyses and provide corresponding insights.

%%%%%%%%%%%%%%%%%%%%%%%%%%%%%%%%%%%%%%%%%%%%%%%%%%%%%%%%%%%%%%%%%%%%%%%%%%
\subsection{HSE06 and Proxy-$U$ Cluster Expansion}
%%%%%%%%%%%%%%%%%%%%%%%%%%%%%%%%%%%%%%%%%%%%%%%%%%%%%%%%%%%%%%%%%%%%%%%%%%

\begin{figure}
\centering
\includegraphics[width=\columnwidth]{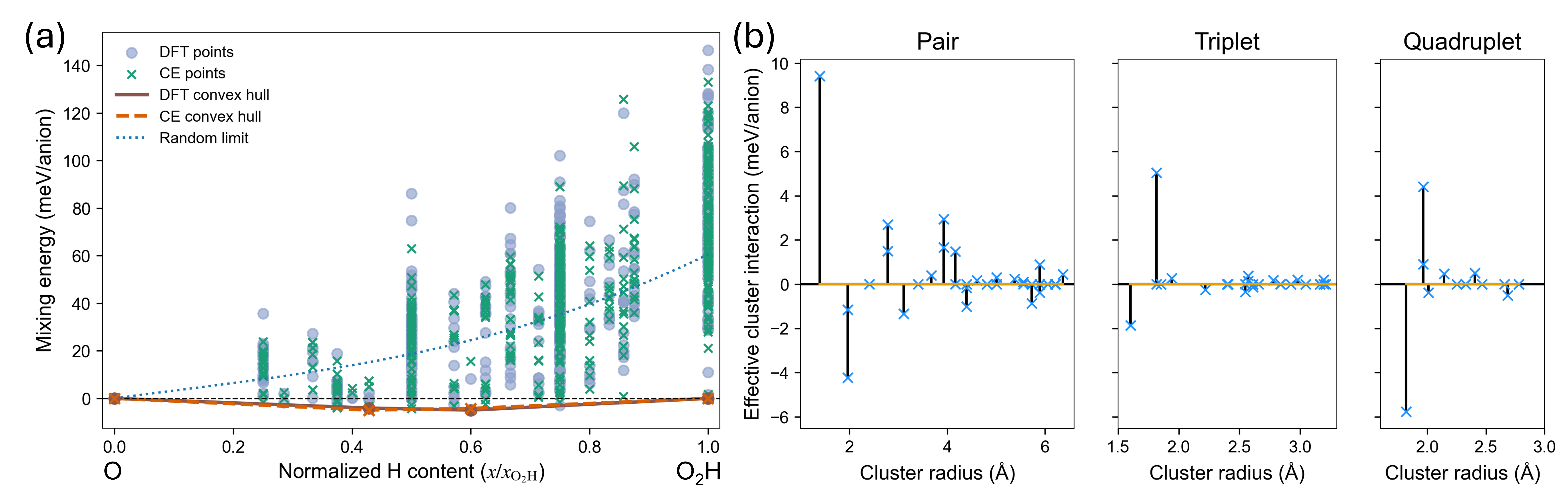}
\caption{\label{fig:proxy-U CE} The proxy-$U$ CE results. (a) DFT convex hull and CE predictions, with the correct orderings on the hull reproduced and a 10-fold CV score of 13.0 meV/anion. The blue dotted line indicates the mixing energy of random configurations at each composition predicted by the CE model. 
(b) Effective cluster interaction (ECI) plot showing that the major contributions arise from pairs, triplets, and quadruplets containing nearest-neighbor (NN) and next-nearest-neighbor (NNN) sites. 
Pair interactions at longer distances decay to zero, and most triplets and quadruplets beyond NNN are negligible. 
}
\end{figure}

After evaluating the sensitivity of DFT settings to ordering energetics and determining the proxy-$U$, we begin the formal construction of the CE described in Figure \ref{fig:framework}.
With a configuration search space of 354,328 supercells and 260 seed structures prepared under the constraints detailed in the Method section, we perform an active-learning scheme to systematically improve the training set and obtain a converged CE model using proxy-$U$.
The resulting proxy-$U$ convex hull comparing CE and DFT predictions, along with the Effective Cluster Interaction (ECI) plot, is shown in Figure \ref{fig:proxy-U CE}.
As expected, new ground states with negative mixing energies appear after the no-$U$ ground states are shifted to higher energies by applying proxy-$U$.
Within our framework, this proxy-$U$ CE is labeled as the intermediate output CE1 and serves as one of the two Hamiltonians used for MC sampling.
Its ECI features are also useful for guiding representative-subset selection.
The next section discusses the ordering features of low-energy configurations in SrTiO$_{3-x}$H$_x$, whereas this section proceeds with the construction of the HSE06 CE.

Since recalculating all 531 structures in the CE1 training set with expensive HSE06 relaxations and energy calculations is computationally demanding, we introduce an additional step to select only a representative subset for HSE06 CE training.
Preserving the logic of the active-learning scheme, we select the subset by including (1) (near) ground-state orderings, in this step, all structures with negative mixing energies in Figure \ref{fig:proxy-U CE}a (12 structures); (2) structures already calculated with HSE06 in Figure \ref{fig:sensitivity} that also appear in the proxy-$U$ training set (16 structures); and (3) structures selected by greedy D-optimal design from the 531 structures in the proxy-$U$ training set. 
In the third step, we add 10 structures each round and test CE fitting using proxy-$U$ energies to determine whether the model acquires enough configurational and energetic information to approximately capture the ECI features in Figure \ref{fig:proxy-U CE}b. 
We finally include 40 structures from this third step of greedy D-optimal selection to satisfy this criteria, and consider key configurations sufficient to determine the general ordering tendencies of SrTiO$_{3-x}$H$_x$ and the most worthwhile for HSE06 calculations. 

\begin{figure}
\centering
\includegraphics[width=\columnwidth]{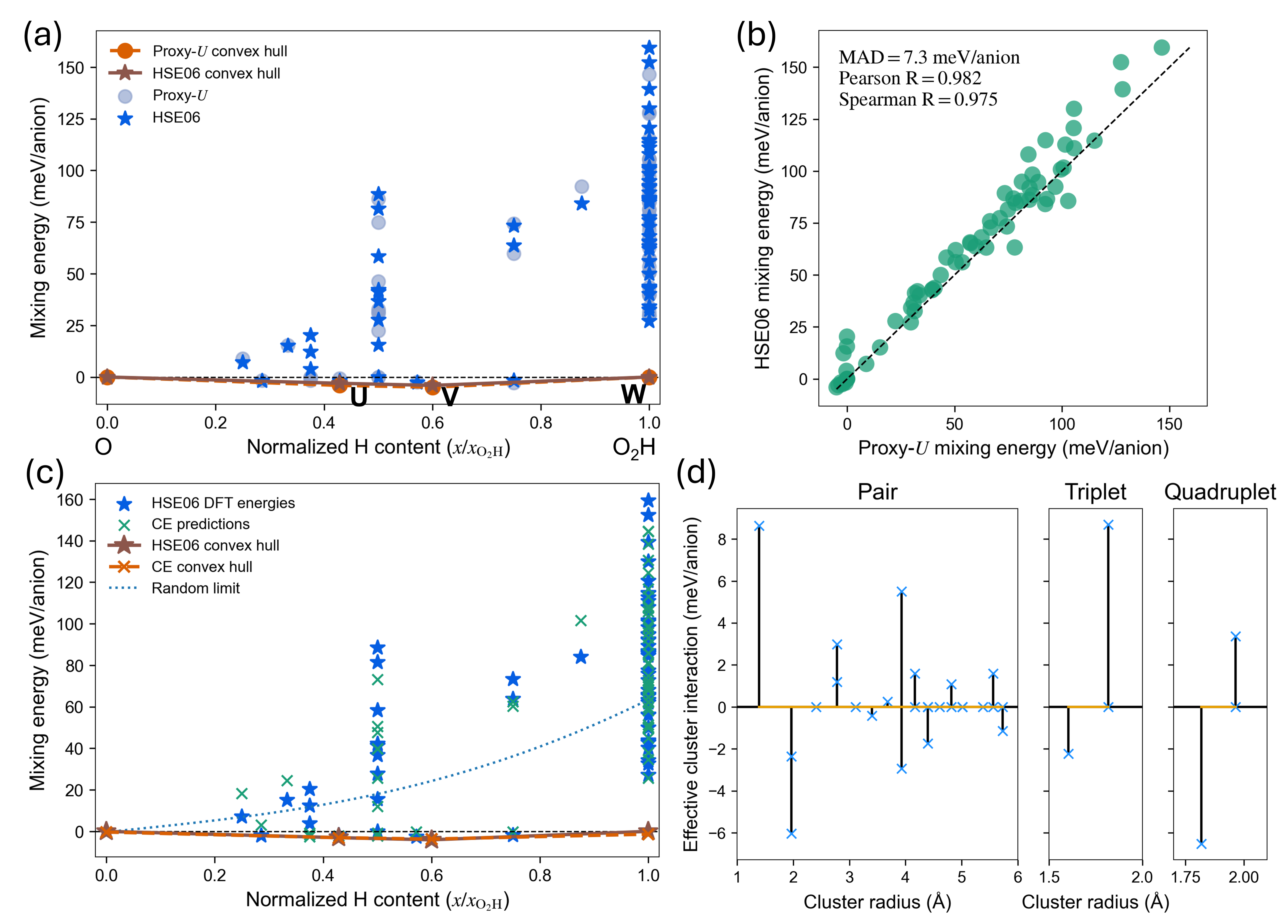}
\caption{\label{fig:HSE06 CE}  HSE06 results, with (a) the convex hull compared with proxy-$U$ data, where both sets of calculations identify the same orderings on the hull at compositions Sr$_7$Ti$_7$O$_{18}$H$_3$, Sr$_5$Ti$_5$O$_{12}$H$_3$,  and SrTiO$_2$H (denoted as U, V, and W. respectively, for discussion in next section), and 
(b) comparison between the mixing energies, showing high Pearson and Spearman correlations and a mean absolute deviation of 7.3 meV/anion. 
(c) HSE06 CE fitting on a minimal training set of 68 key configurations, with the correct orderings on the hull reproduced and a 5-fold CV score of 28.1 meV/anion.
(d) ECI plot showing trends similar to those in the proxy-$U$ version, but with reduced cluster cutoffs.
}
\end{figure}

These 68 key configurations, with HSE06 relaxations and energy calculations, are then used to build the final output CE2 and are also directly compared with the proxy-$U$ results, as presented in Figure \ref{fig:HSE06 CE}. 
We observe that, across this entire HSE06 convex hull, which includes an extensive set of structures beyond those used in the proxy-$U$ fitting step of the preliminary stage, the low-cost proxy-$U$ predictions still agree well with the HSE06 results, despite being obtained from two fundamentally different functionals.
Among these configurations, HSE06 and proxy-$U$ agree on the ground-state orderings on the hull.
We consider HSE06 energetics to be physically more accurate, but proxy-$U$ results, which show a mean absolute deviation in mixing energy of 7.3 meV/anion relative to HSE06, may be sufficient for more extensive and resource-intensive future work, such as large-scale MLIP training, provided they are used with caution and cross-checked against HSE06 results.
During HSE06 CE fitting with a limited amount of data, we first reduce the cluster cutoffs based on what we learned from the proxy-$U$ CE, retaining only triplets and quadruplets containing up to NNN sites. 
We then perform the hyperparameter optimizations detailed in the Method section to obtain the optimal fit, which produces the final output CE2 trained on pure HSE06 energetics. 
Within these key configurations, the proxy-$U$ CE1 and HSE06 CE2 both reproduce the same orderings on the hull and show similar ECI trends for the important short-range clusters.
We note the caveat that, when new ground-state predictions are performed using CE2, additional nearly degenerate configurations appear and deepen the hull slightly. 
These candidates show the same overall ordering patterns with only minor variations, which may reflect the limited fitting resolution of CE2.
We therefore do not expect them to alter the ordering analysis based on this CE2, and do not pursue them with further HSE06 calculations.
We also note that, although CE1 exhibits a lower CV score and the proxy-$U$ energetics are reasonably close to HSE06 on average, this observation does not by itself establish that CE1 is more reliable than the directly trained HSE06 model CE2 for extrapolating large-supercell energetics and finite-temperature properties through MC sampling, as such predictions can depend not only on average fitting error.
A more complete assessment of which model is preferable under practical resource constraints remains an interesting subject for future study, but is not explored further in this work.
In the final section on anion ordering tendency, we present results from both CE models and reveal the strong O$_4$H$_2$ cis preference in SrTiO$_{3-x}$H$_x$ suggested by both CE models.

%%%%%%%%%%%%%%%%%%%%%%%%%%%%%%%%%%%%%%%%%%%%%%%%%%%%%%%%%%%%%%%%%%%%%%%%%%
\subsection{Anion Ordering Tendency: O$_4$H$_2$ cis Stabilization}
%%%%%%%%%%%%%%%%%%%%%%%%%%%%%%%%%%%%%%%%%%%%%%%%%%%%%%%%%%%%%%%%%%%%%%%%%%

\begin{figure}
\centering
\includegraphics[width=\columnwidth]{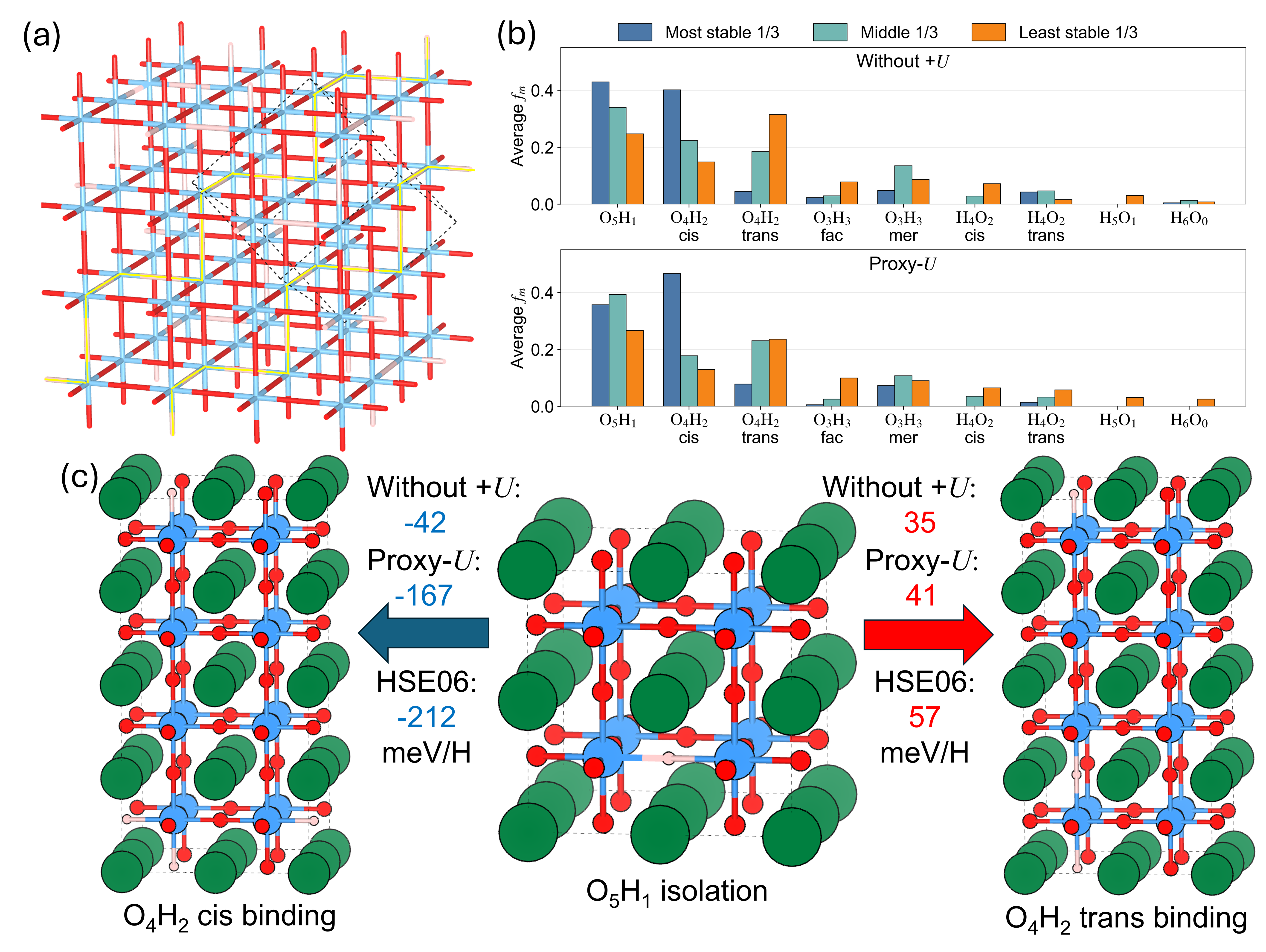}
\caption{\label{fig:Anion_ordering} Anion ordering analysis. (a) The Ti-H and Ti-O bonding framework of ground-state Sr$_5$Ti$_5$O$_{12}$H$_3$ (structure V in Figure \ref{fig:HSE06 CE}a). The blue cubic corners are the Ti positions, the red sections represent O, and the pink sections represent H, illustrating the Ti-H and Ti-O bonds around Ti. 
The 90$^\circ$ Ti-H bonds repeatedly shown in the plot correspond to the O$_4$H$_2$ cis configuration.
These cis chains, highlighted with thin yellow lines, can extend across different cubic $\{100\}$ planes and propagate along $\langle 111 \rangle$ directions, which we denote as the cis-stair pattern. The dashed line indicates the periodic unit cell of this Sr$_5$Ti$_5$O$_{12}$H$_3$ structure. 
(b) The histograms of the average for the octahedral fraction (defined in Equation \ref{eq:Ti-centric}) for structures classified into terciles according to their DFT convex hull energies, calculated from both proxy-$U$ dataset and the same set of structures recalculated without +$U$. 
O$_4$H$_2$ cis is identified as the dominant octahedral configuration in low-energy structures, especially when the over-delocalization error is corrected.
(c) Direct evaluation of cis and trans H$_2$ binding energies by calculating the energy difference between $2 \times 2 \times 2$ (O$_5$H$_1$ isolation) and $2 \times 2 \times 4$ supercells (O$_4$H$_2$ pair binding) at the dilute composition Sr$_8$Ti$_8$O$_{23}$H.
Under both proxy-$U$ and HSE06, the stabilization from O$_4$H$_2$ cis is approximately -200 meV/H and might be non-negligible even at dilute composition.
}
\end{figure}

In this final section, we present computational evidence from multiple perspectives supporting the prevalent O$_4$H$_2$ cis preference in SrTiO$_{3-x}$H$_x$.
We first discuss octahedral configurations for the ground-state orderings (U, V, and W in Figure \ref{fig:HSE06 CE}a).
Figure \ref{fig:Anion_ordering}a illustrates the Ti-H and Ti-O bonding framework of the ordered structure V of Sr$_5$Ti$_5$O$_{12}$H$_3$, with all H-containing O$_{6-x}$H$_x$ octahedra surrounding Ti adopting the O$_4$H$_2$ cis configuration.
All U, V, and W structures exhibit similar O$_4$H$_2$ cis patterns through corner-sharing octahedra arrangements across their supercells, which we denote as the cis-stair pattern hereafter. 
This pattern differs from the arrangement in which all cis octahedra lie in the cubic \{100\} planes, as described in previous perovskite oxynitride literature.\cite{yang2011anion}
The U and W structures exhibit the similar cis-stair pattern propagating along $\langle 111 \rangle$ directions but with different spacing between the cis-stair patterns. 
The cis-stair in W structure of SrTiO$_2$H stacks closely that all of the octahedra in it are completely O$_4$H$_2$ cis, matching the O to H ratio of 2 to 1 both locally within the octahedra and globally.
However, the order-disorder temperatures are expected, based on the MC sampling used later in this section, to be below room temperature, especially in the region of limited hydride incorporation ($x<0.5$) in synthesized SrTiO$_{3-x}$H$_x$. 
Moreover, many other cis-type patterns are nearly degenerate, so this cis-stair pattern is likely hypothetical and difficult to observe experimentally.

After examining the specific cases of the orderings on the hull, we provide a statistical overview (Figure \ref{fig:Anion_ordering}b) of how DFT energetic stability is related to octahedral configurations, based on the CE1 training set of 531 structures with diverse octahedral environments.
We define the fraction $f_m$ of octahedral configuration $m$ in a given structure as:
\begin{equation}
f_m\label{eq:Ti-centric}
=
\frac{n_m}{\sum\limits_{k \in S} n_k}
\end{equation}
where $n_m$ is the number of octahedral configuration $m$ within given structure, 
and $S$ denotes the set of H-containing octahedral (excluding pure O octahedra) included in the analysis.
With proxy-$U$, structures with a high fraction of O$_4$H$_2$ cis dominate the most stable one-third of the 531 structures in the CE1 training set.
When the over-delocalization treatment is turned off, the preference shifts away from O$_4$H$_2$ cis toward O$_5$H$_1$, while O$_4$H$_2$ trans consistently appears most frequently among the least stable configurations.
As expected from the high correlations between HSE06 and proxy-$U$ energetics (Figure \ref{fig:HSE06 CE}b), the similar analysis of the HSE06 training set, although based on only 68 key configurations, shows the same trend of O$_4$H$_2$ cis stabilization, with the corresponding histogram provided in the Supporting Information.
While this statistic provides a qualitative hint of how the energy varies with the fraction of different octahedral configurations in each structure, we next present direct energy evaluations of H$_2$ cis-type binding in dilute anion mixing supercells compared with the isolated H model commonly used in the literature.\cite{liu2019highly, granhed2019band}

As demonstrated in Figure \ref{fig:Anion_ordering}c, anion ordering remains energetically relevant even at the dilute composition Sr$_8$Ti$_8$O$_{23}$H. 
Consistent with the trend observed in the previous histogram, cis-type H$_2$ pair formation provides additional stabilization relative to the reference case of isolated H at the same composition.
This stabilization is nearly 200 meV/H under proxy-$U$ and exceeds 200 meV/H with HSE06, but is less pronounced with plain PBE without +$U$.
In contrast, trans pairing is always unstable relative to isolated H, which explains why structures containing this configuration are commonly among the least stable.
For perovskite oxyhydride studies employing DFT+$U$, the magnitude of cis-type stabilization, if taken into account, may be comparable to or even exceed other energetic quantities computed within the isolated-H picture, such as migration barriers and polaron formation energies.\cite{liu2019highly, granhed2019band} 
These results suggest that anion ordering tendencies should be considered in the structural model, rather than routinely assuming an isolated-H configuration in the supercell across both dilute and non-dilute composition ranges.

\begin{figure}
\centering
\includegraphics[width=\columnwidth]{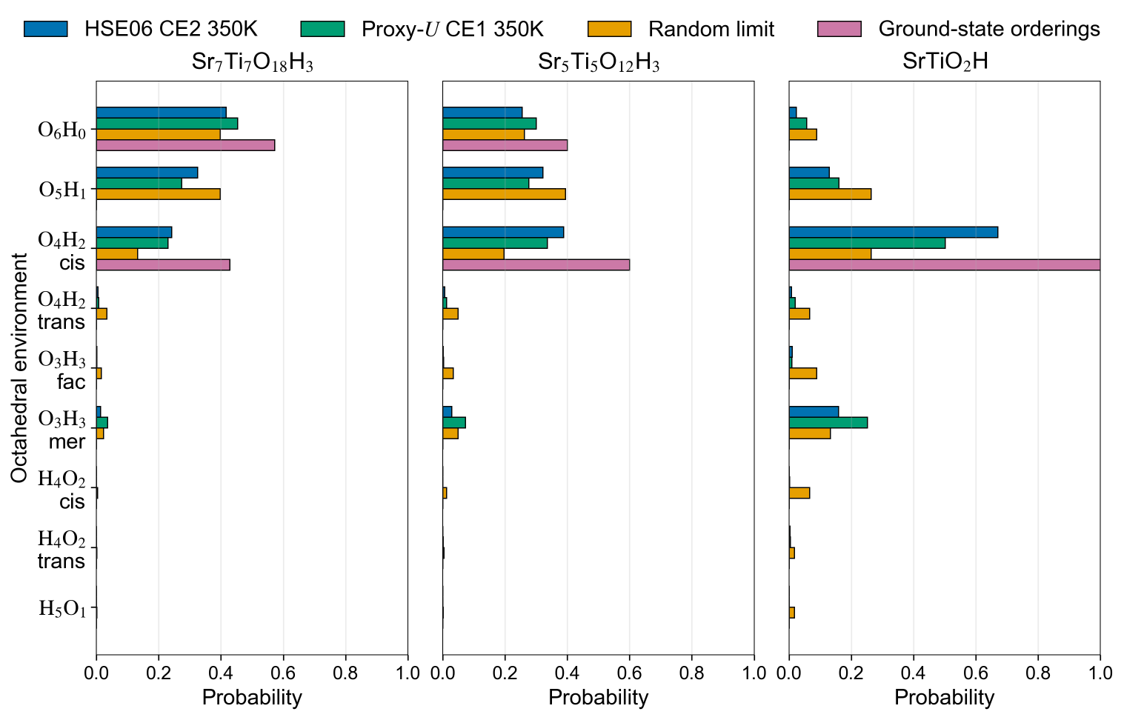} 
\caption{\label{fig:MC} MC sampling octahedral probabilities from both proxy-$U$ CE1 and HSE06 CE2 at 350 K, above the disordering temperatures, with the random limit and ground-state values also plotted for comparison.
Ground-state orderings are those labeled U, V, and W in Figure \ref{fig:HSE06 CE}a.
In SrTiO$_{3-x}$H$_x$, MC results for the disordered states from both CE models show similar trends, shifting away from the random limit toward the O$_4$H$_2$-rich ground states.
}
\end{figure}

Finally, we report MC sampling results for octahedral probabilities in disordered SrTiO$_{3-x}$H$_x$ using both the proxy-$U$ CE1 and HSE06 CE2 models.
As shown in Figure \ref{fig:MC}, octahedral probabilities in disordered states of Sr$_7$Ti$_7$O$_{18}$H$_3$, Sr$_5$Ti$_5$O$_{12}$H$_3$, and SrTiO$_2$H still exhibit SRO with large deviations from the random distribution value.
At SrTiO$_2$H, consistent with previous observations, the disordered state exhibit a strong preference for forming local O$_4$H$_2$ cis configurations, where O$_5$H$_1$ formation is accompanied by O$_3$H$_3$ octahedra, mostly in the mer configuration, to maintain the global O:H ratio of 2:1.
However, even with less H incorporation (around Sr$_7$Ti$_7$O$_{18}$H$_3$ to Sr$_5$Ti$_5$O$_{12}$H$_3$, closer to the maximum H level reported in oxyhydrides),\cite{kobayashi2012oxyhydride, sakaguchi2012oxyhydrides} the O$_4$H$_2$ cis population in the disordered states remains nearly comparable to that of O$_5$H$_1$. 
HSE06 and proxy-$U$ predictions are qualitatively consistent, with HSE06 showing a stronger preference for cis configurations than proxy-$U$.
This ordering preference might fundamentally affect the modeling of other material properties in SrTiO$_{3-x}$H$_x$.
For example, for H migration, if moving H from one site to another changes the ordering pattern from a stable cis pair to a less favorable end configuration, it is worth questioning whether the overall diffusion event can still be properly described by an isolated-H picture, in which H moves between two equivalent sites and only the migration barrier of the intermediate state is considered.
Taking a step back, even without considering cis ordering preferences and focusing only on non-dilute effects through the random limit at these compositions, the isolated-H picture remains questionable. 
Assuming isolated H already imposes a particular ordering tendency, namely that H atoms prefer to remain far apart from one another beyond nearest neighbors, and therefore does not provide a good representation of a random configuration or of disordered states in general.
In summary, we show in this section that anion-ordering behavior and realistic non-dilute cells can play a nontrivial role in the accurate simulation of oxyhydrides.

%%%%%%%%%%%%%%%%%%%%%%%%%%%%%%%%%%%%%%%%%%%%%%%%%%%%%%%%%%%%%%%%%%%%%%%%%%
\subsection{Conclusion}
%%%%%%%%%%%%%%%%%%%%%%%%%%%%%%%%%%%%%%%%%%%%%%%%%%%%%%%%%%%%%%%%%%%%%%%%%%
In this study, we investigate anion ordering in SrTiO$_{3-x}$H$_x$ with a cluster expansion framework based on HSE06 and proxy-$U$ energetics. 
We first reveal that anion ordering stability is sensitive to the choice of DFT+$U$, an aspect that has been underexplored in conventional CE studies, and extend CE modeling to incorporate hybrid-functional-level accuracy in the training set energetics.
The framework, designed to efficiently select key configurations that span the configurational space while minimizing the number of expensive HSE06 calculations, may also be applicable to other systems that require hybrid-functional DFT treatment.
Our results show that SrTiO$_{3-x}$H$_x$ strongly favors the O$_4$H$_2$ cis configuration, suggesting that this ordering tendency may play a non-negligible role across a broad composition range in simulations of material properties in this system.

\clearpage
\begin{acknowledgement}
This research was supported as part of the Hydrogen in Energy and Information Sciences (HEISs), an Energy Frontier Research Center funded by the U.S. Department of Energy (DOE), Office of Science, Basic Energy Sciences (BES), under award No. DE-SC0023450.
The authors acknowledge resources from the Northwestern Quest high-performance computing facility, which is jointly supported by the Office of the Provost, the Office for Research, and Northwestern University Information Technology.
HSE06 calculations were performed using computational resources at the National Energy Research Scientific Computing Center (NERSC), a Department of Energy User Facility using NERSC award BES-ERCAP0032806.
\end{acknowledgement}

\section*{AI Technology Usage Disclosure}
AI tools were used responsibly for language refinement and coding assistance. 
ChatGPT and Grammarly Premium were used during manuscript preparation to assist with grammar and language clarity. 
Suggestions were evaluated and applied only when the authors considered them beneficial and accurate.
AI coding assistants, primarily Codex within the Visual Studio Code integrated development environment, were used to assist with drafting, debugging, and refining scientific code. 
All content was thoroughly reviewed and revised to ensure accuracy, and approved by both authors.
The authors retain full responsibility for all scientific content, analyses, and conclusions presented in this work.

\begin{suppinfo}
Supplementary histogram of the average octahedral fractions for 68 key configurations classified into terciles based on their HSE06 convex-hull energies.
\end{suppinfo}

\clearpage
\bibliography{achemso-demo}

\clearpage

\section*{Supporting Information}
\begin{figure}
\centering
\includegraphics[width=\columnwidth]{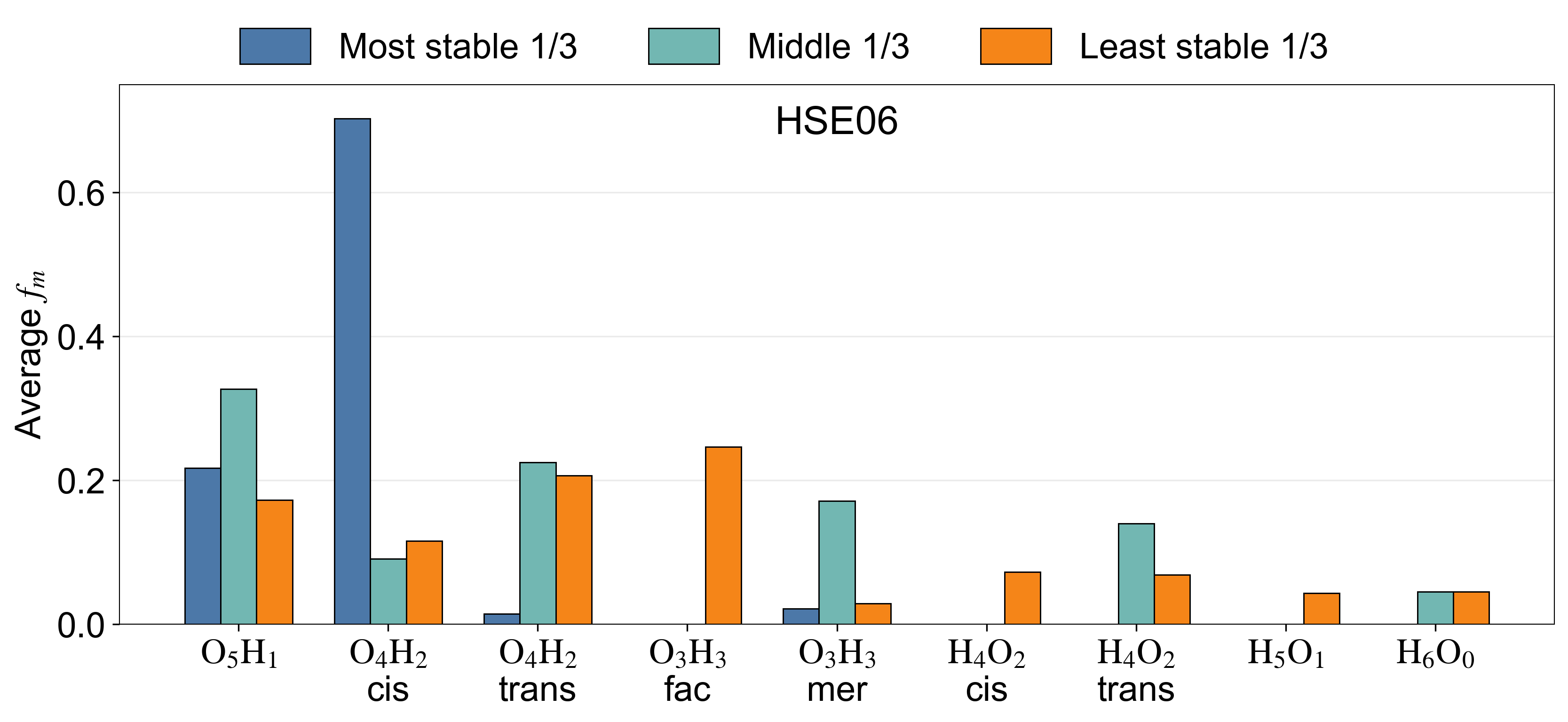} 
\caption{\label{fig:SI_hist} The histogram of the average for the octahedral fraction for structures classified into terciles according to their HSE06 convex hull energies.
O$_4$H$_2$ cis is the dominant octahedral configuration in the most stable one-third of the structures.
Please note that this histogram is based on 68 key configurations and therefore cannot be directly compared with Figure \ref{fig:Anion_ordering}b, which is derived from the 531 structures in the CE1 training set.
}
\end{figure}

\end{document}